\documentclass[secnumarabic,aps,prd,floatfix,nofootinbib,showkeys,twocolumn,showpacs,10pt,superscriptaddress] 
{revtex4-1}

\usepackage{times}
\usepackage{amsfonts}
\usepackage{amsmath}
\usepackage{amssymb}
\usepackage{natbib}
\usepackage{graphicx}
\usepackage{enumitem}
\usepackage{xcolor}
\usepackage[colorlinks=true,linkcolor=red, citecolor=blue]{hyperref}
\usepackage[outdir=./]{epstopdf}
\usepackage[normalem]{ulem}
\usepackage{amsmath}

\def\aap{A\&A}
\def\aj{AJ}
\def\apj{ApJ}
\def\apjl{ApJL}
\def\apjs{ApJS}

\def\jcap{JCAP}
\def\mnras{MNRAS}
\def\na{New Astronomy}
\def\nar{New Astronomy Reviews}

\begin{document}
\definecolor{orange}{rgb}{0.9,0.45,0}
\def\CovDev{D}
\def\Res{{\mathcal R}}
\def\Gammaflat{\hat \Gamma}
\def\metricflat{\hat \gamma}
\def\Dflat{\hat {\mathcal D}}
\def\part_n{\partial_\perp}
%
\def\Lie{\mathcal{L}}
\def\A{\mathcal{X}}
\def\Aphi{\A_{\phi}}
\def\hAphi{\hat{\A}_{\phi}}
\def\E{\mathcal{E}}
\def\Ham{\mathcal{H}}
\def\M{\mathcal{M}}
\def\R{\mathcal{R}}
\def\p{\partial}
\def\hg{\hat{\gamma}}
\def\hA{\hat{A}}
\def\hD{\hat{D}}
\def\hE{\hat{E}}
\def\hR{\hat{R}}
\def\hcA{\hat{\mathcal{A}}}
\def\hDelt{\hat{\triangle}}
\def\na{\nabla}
\def\dif{{\rm{d}}}
\def\non{\nonumber}
\newcommand{\erf}{\textrm{erf}}
\newcommand{\saeed}[1]{\textcolor{blue}{SF: #1}} 
%
\renewcommand{\t}{\times}
\long\def\symbolfootnote[#1]#2{\begingroup%
\def\thefootnote{\fnsymbol{footnote}}\footnote[#1]{#2}\endgroup}
\title{High-redshift Galaxies from JWST Observations in More Realistic Dark Matter Halo Models} 

\author{Saeed Fakhry}
\email{s\_fakhry@kntu.ac.ir}
\affiliation{Department of Physics, K.N. Toosi University of Technology, P.O. Box 15875-4416, Tehran, Iran}

\author{Reyhaneh Vojoudi Salmani} 
\email{r.s.vojoudi@gmail.com}
\affiliation{Department of Physics, K.N. Toosi University of Technology, P.O. Box 15875-4416, Tehran, Iran}

\author{Javad T. Firouzjaee} 
\email{firouzjaee@kntu.ac.ir}
\affiliation{Department of Physics, K.N. Toosi University of Technology, P.O. Box 15875-4416, Tehran, Iran}
\date{\today}

\begin{abstract} 
\noindent
The James Webb Space Telescope (JWST) has unveiled unexpectedly massive galaxy candidates at high redshifts, challenging standard $\Lambda$CDM cosmological predictions. In this work, we study the predictions of more realistic dark matter halo models combined with modified matter power spectra for interpreting JWST observations of high-redshift galaxies. We employ three halo mass functions: the conventional Sheth-Tormen (ST) model and two physically motivated alternatives introduced by Del Popolo (DP1 and DP2) that incorporate angular momentum, dynamical friction, and cosmological constant effects. These are coupled with parametrically modified power spectra featuring small-scale enhancements characterized by spectral indices and characteristic scales, motivated by cosmological scenarios. Our analysis of cumulative stellar mass densities at $z \simeq 8$--$10$ reveals that the standard ST mass function systematically underpredicts JWST observations, achieving marginal consistency only with high star formation efficiencies. In contrast, the DP1 and DP2 models demonstrate significantly improved agreement with observations even within standard $\Lambda$CDM, with statistical consistency within $1$--$2\sigma$ for moderate star formation efficiencies. When combined with modified power spectra, these refined halo models achieve suitable agreement with JWST data across broad parameter ranges, particularly for steeper spectral indices that amplify high-mass halo formation. Crucially, we find that moderate star formation efficiencies coupled with small-scale power enhancements provide robust reconciliation between theory and observations, eliminating the need for extreme astrophysical assumptions. Our results demonstrate that incorporating realistic halo collapse physics, often neglected in standard analyses, can substantially alleviate apparent tensions between JWST observations and $\Lambda$CDM predictions, highlighting the critical importance of small-scale structure formation physics in early cosmic epochs.
\end{abstract}


\maketitle
\vspace{0.8cm}

\section{Introduction} 
The Lambda cold dark matter ($\Lambda$CDM) model has served as the standard cosmological paradigm since the discovery of cosmic acceleration through supernova observations, providing a robust framework that successfully explains the majority of large-scale structure observations \citep{1998AJ....116.1009R}. Despite this success, the increasing precision of cosmological measurements has revealed several persistent tensions that challenge the internal consistency of this model \citep{2016IJMPD..2530007B, 2021JHEAp..32...28A, 2021CQGra..38o3001D, 2022JHEAp..34...49A, 2022NewAR..9501659P}. Most notably, independent determinations of the Hubble constant ($H_{0}$) from early-Universe probes such as the cosmic microwave background (CMB) and baryon acoustic oscillations (BAO) yield values that are statistically incompatible with late-time measurements from supernovae and gravitational lensing, with discrepancies now exceeding $5\sigma$ significance \citep{2021MNRAS.502.2065D}. Similarly, the matter clustering amplitude ($\sigma_{8}$) exhibits tension between early and late-time measurements, while the cosmological lithium problem persists, with Big Bang nucleosynthesis predictions underestimating primordial lithium abundances by up to $4\sigma$ tensions \citep{2011ARNPS..61...47F}. Additional challenges arise from dynamical anomalies in structure formation, such as the unexpectedly high collision velocity observed in the \textit{El Gordo} galaxy cluster \citep{2021MNRAS.500.5249A}, and from conceptual issues surrounding the inflationary paradigm, including the vast landscape of solutions \citep{2013JCAP...11..040M}, and the problem of past-incomplete geodesics \citep{2003PhRvL..90o1301B}. 

The evaluation of the $\Lambda$CDM model at high redshifts ($z>7$) remains challenging due to the limited observational reach of current primary cosmological probes, particularly BAO and Type Ia supernovae. The absence of reliable high-redshift observations via the mentioned probes constrains our ability to rigorously test $\Lambda$CDM predictions during the early stages of cosmic evolution. Fortunately, recent transformative observations from the James Webb Space Telescope (JWST) offer a valuable opportunity to critically assess the $\Lambda$CDM model at high redshifts. Particularly, the JWST has recently detected galaxy candidates in high redshifts exhibiting unexpectedly high stellar masses \citep{2022ApJ...938L..15C, 2022ApJ...940L..14N, 2023Natur.616..266L, 2023ApJS..265....5H, 2023NatAs...7..622C, 2024ApJ...964...71H, 2023NatAs...7..611R}. The presence of these massive galaxies at such early cosmic times poses a significant challenge to the predictions of the $\Lambda$CDM framework \citep{2023NatAs...7..731B, 2023MNRAS.518.2511L}. Hence, the substantial discrepancies between observational constraints and $\Lambda$CDM predictions highlight the critical need for either framework refinement, incorporation of neglected physical mechanisms, or development of an alternative theoretical paradigm to address these tensions.

Several works have attempted to reconcile $\Lambda$CDM cosmology with JWST observations of high-redshift galaxy candidates. An initial explanation is an enhanced star formation efficiency, particularly at high redshifts. Hydrodynamical simulations suggest that this efficiency is not constant but may increase markedly in earlier cosmic epochs, see, e.g., \citep{2023A&A...677L...4P, 2024A&A...689A.244C}, potentially driven by feedback-free starburst events, which are more frequent in high-density, low-metallicity environments, e.g., \citep{2023MNRAS.523.3201D, 2025ApJ...980..138H}. Moreover, the absence of ultraviolet (UV) background suppression during the pre-reionization era could further elevate star formation rates, e.g., \citep{2023ApJS..265....5H}. An alternative interpretation focuses on the intrinsic UV luminosity of high-redshift galaxies. Efficient dust ejection during early galaxy assembly may render dust attenuation negligible, thereby increasing apparent brightness \citep{2023MNRAS.522.3986F, 2024PDU....4401496I}. Active galactic nuclei (AGNs) may also contribute to UV luminosity, as indicated by their detection in systems such as \text{CEERS\_1019} and GN-z11, e.g., \citep{2023A&A...677A..88B, 2023ApJ...952...74T, 2023ApJ...953L..29L, 2023ApJ...959...39H}. However, the lack of AGN signatures in many galaxies suggests that stellar processes may dominate their emission \citep{2023ApJ...959...39H}. Another hypothesis challenges the universality of the classical Salpeter Initial Mass Function (IMF) and its variants \citep{1955ApJ...121..161S, 2019NatAs...3..482K, 2023ApJ...951L..40S}. A non-universal, top-heavy IMF favoring massive stars could account for the excess of bright galaxies at high redshift, e.g., \citep{2023ApJ...951L..40S, 2024MNRAS.534..523C}. Yet, this scenario faces challenges due to limited observational support and the counteracting influence of stronger stellar feedback, e.g., \citep{2024A&A...686A.138C}.

In addition to the mentioned efforts, other alternative approaches have been proposed and investigated to address the tension between the JWST observations of high-redshift galaxies and the predictions of the  $\Lambda$CDM model. These include models incorporating a dynamical dark energy component characterized by a positive energy density and a negative cosmological constant \citep{2023JCAP...10..072A, 2024JCAP...07..072M}, modifications to the power spectrum to account for small-scale statistical features \citep{2023MNRAS.526L..63P, 2023PhRvD.107d3502H, 2024PhLB..85839062B}, scenarios involving the dynamics of primordial black holes (PBHs) \citep{2023arXiv230617577H, 2024SCPMA..6709512Y}, early dark energy scenarios \citep{2024MNRAS.533.3923S, 2024JCAP...05..097F, 2025PhRvD.111b3519J}, non-standard dark matter models \citep{2024MNRAS.534.2848D}, and extensions to the $\Lambda$CDM model through modified theories of gravity \citep{2024arXiv241203534M, 2025arXiv250111103S}.

Beyond these methodological frameworks, an additional significant avenue can involve the incorporation of more physically realistic dark matter halo models. Such models assume particular relevance given that dark matter constitutes the dominant component of galactic mass distributions and may serve as a fundamental driver in the formation and evolutionary processes of intraglactic structures. In this regard, one of the most important factors in describing dark matter halo models is the halo mass function, which leads to the classification of dark matter halos in terms of their mass. The conventional Press-Schechter (PS) mass function \citep{1974ApJ...187..425P}, while foundational, exhibits some deviations from observational data and $N$-body simulations, introducing systematic uncertainties in cosmological analyses. The observed discrepancy can be attributed to specific geometrical and physical factors that are neglected within the PS formalism, despite their critical importance in accurately predicting halo mass functions. The Sheth-Tormen (ST) formalism constitutes a suitable refinement of this theoretical approach, employing ellipsoidal-collapse models with dynamical density barrier to provide improved consistency with high-resolution numerical simulations when compared to the original PS formulation \citep{1999MNRAS.308..119S, 2001MNRAS.323....1S}. 

Although simulation-based mass functions like ST offer valuable approximations, their black-box nature obscures individual physical contributions to halo mass function morphology, particularly at high redshift \citep{2007MNRAS.374....2R, 2007ApJ...671.1160L}.. Beyond geometric factors during halo virialization, various physical processes substantially affect overdensity collapse and the resulting mass function. Integrating these core physical elements into theoretical models is vital for capturing fundamental physics controlling halo formation throughout cosmic structure development. This methodology enables dynamic collapse criteria that adapt with relevant physical variables, producing more realistic halo collapse modeling. One such theoretical refinement can incorporate the effects of angular momentum, dynamical friction, and the cosmological constant into the halo mass function \citep{2006ApJ...637...12D, 2017JCAP...03..032D}, significantly mitigating observed discrepancies, especially at the high-mass end and high-redshift regime. Incorporating such effective geometrical and physical factors into the halo mass function formalism may enhance the precision of halo model forecasts when compared to observational data, see, e.g., \citep{2021PhRvD.103l3014F, 2022PhRvD.105d3525F, 2022ApJ...941...36F, 2023PDU....4101244F, 2023PhRvD.107f3507F, 2023ApJ...947...46F, 2023arXiv230811049F, 2024PDU....4601544F, 2024arXiv240115171F, 2024PDU....4601712T, 2025arXiv250200914F}.

In this study, we investigate galaxy candidates observed at high redshifts by the JWST within the framework of more realistic dark matter halo models. The structure of the paper is as follows. In Sec.\,\ref{sec:ii}, we provide an analytical formulation of the modified power spectrum, incorporating plausible scenarios for small-scale modifications. In Sec.\,\ref{sec:iii}, we introduce more realistic halo mass functions that include both geometrical and physical corrections. In Sec.\,\ref{sec:iv}, we develop a theoretical model of star formation and compare the resulting predictions from provided dark matter halo models with JWST observations of high-redshift galaxy candidates. Finally, in Sec.\,\ref{sec:v}, we discuss our results and explore their broader implications.
\section{Modified Matter Power Spectrum}\label{sec:ii}
The matter power spectrum plays a central role in understanding the formation of cosmic structures. It quantifies the distribution of matter density fluctuations across different scales and serves as a cornerstone for predicting the abundance and clustering of dark matter halos. In the standard $\Lambda$CDM model, the matter power spectrum is well-constrained on large scales ($k\lesssim 3\,h\,{\rm Mpc^{-1}}$) by observations of the galaxy clustering, Lyman-$\alpha$ forest and the UV luminosity function \citep{2022ApJ...928L..20S, 2019MNRAS.489.2247C, 2010MNRAS.404...60R}

However, at smaller scales, considerable deviations from the standard $\Lambda$CDM model may manifest, exerting a pronounced influence on the statistical distribution of massive dark matter halos. The fundamental physics underlying these deviations stems from the fact that alternative dark matter models or modified gravity theories can substantially alter the matter power spectrum at scales where nonlinear structure formation becomes dominant \citep{2012MNRAS.421...50V, 2019NatAs...3..945A}. Moreover, such modifications may emerge from inflationary models that incorporate a scale-dependent spectral index or introduce localized spectral features \citep{2017PDU....17...38G, 2024MNRAS.530.1424L}. A blue-tilted power spectrum amplifies small-scale density fluctuations, increasing the abundance of massive dark matter halos at high redshift \citep{2024ApJ...963....2H}. This enhanced halo formation could explain the unexpectedly luminous and high-redshift massive galaxies by JWST.

As mentioned above, alternative explanations invoke non-standard dark matter physics that modify the small-scale power spectrum. Fuzzy dark matter suppresses structure below the Jeans scale through quantum pressure effects, while axion dark matter enhances small-scale power via dense miniclusters formed during the QCD phase transition \citep{2023PhRvD.107d3502H, 2024PhLB..85839062B}. PBHs similarly boost small-scale fluctuations through compensated isocurvature perturbations and gravitational clustering during radiation domination, potentially explaining enhanced early star formation rates \citep{2024PhRvD.110j3540H}. In this regard, the halo mass function is directly influenced by this scale-dependent modification in clustering properties. The high-mass tail shows the most significant effects due to its exponential sensitivity to the underlying power spectrum. This enhanced sensitivity may lead to predictions that align with JWST observations of high-redshift galaxies.

The modifications to the matter power spectrum can be expressed mathematically as \citep{2023PhRvD.107d3502H}:
\begin{equation}
\mathcal{P}_{\rm Mod}(k) = \mathcal{P}_{\Lambda\rm CDM}(k)+ \mathcal{P}_{\Lambda{\rm CDM}}(k_{c})\left(\dfrac{k}{k_{c}}\right)^{n},
\label{Eq1}
\end{equation}
where $\mathcal{P}_{\Lambda\rm CDM}(k)$ represents the standard $\Lambda$CDM power spectrum, $k_c > 3\, h\,{\rm Mpc^{-1}}$ defines the characteristic scale above which the power-law regime dominates the spectrum, and $n$ denotes the spectral index characterizing the slope of the power-law tail. A cutoff is introduced in the power-law regime of the power spectrum at a scale $k_{\text{cut}} > k_c$. As discussed later, the results are largely insensitive to the specific value of this cutoff, provided that $k_{\text{cut}} \gtrsim 30\,h\,\mathrm{Mpc}^{-1}$. The $\Lambda$CDM matter power spectrum is derived from a nearly scale-invariant power spectrum with a spectral index $n_s = 0.965$, in combination with the transfer function fitted in \citep{1998ApJ...496..605E}.

\begin{figure}
\centering
\includegraphics[width=1\linewidth]{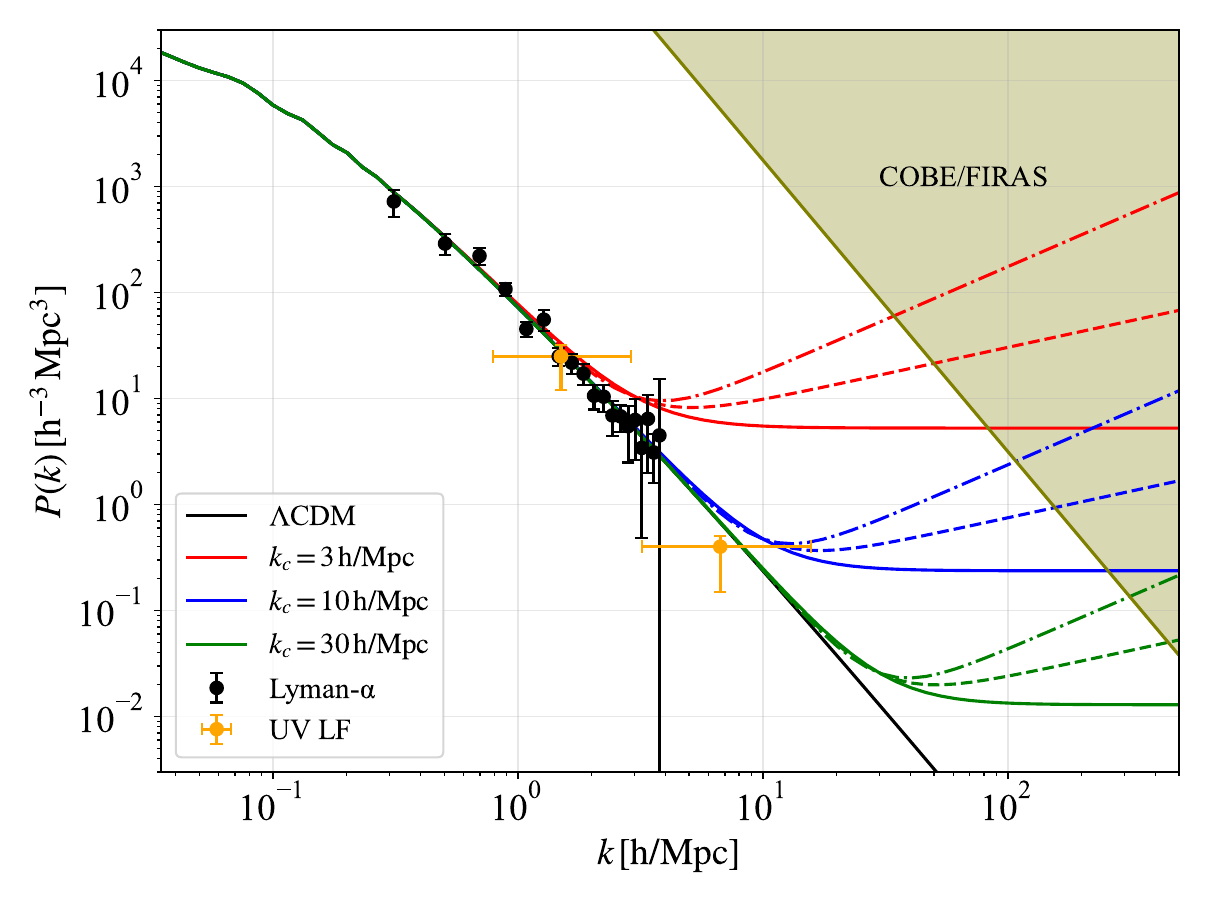}
\caption{The modified matter power spectra derived from Eq.\,\eqref{Eq1} are presented for three values of the characteristic scale $,k_{c} = 3, 10$, and $30\, h\,{\rm Mpc^{-1}}$, in comparison with the matter power spectrum of the standard $\Lambda$CDM model. Three distinct values for the power-law spectral index, $n = 0, 0.5$, and $1$, are considered, and are represented by solid, dashed, and dot-dashed lines, respectively. The black and orange data points, along with their respective error bars, represent Lyman-$\alpha$ measurements \citep{2019MNRAS.489.2247C}, and HST measurements of the UV luminosity function \citep{2022ApJ...928L..20S}, respectively. The olive shaded region denotes the maximum cut-off scale for $k^4$ growth in the adiabatic curvature power spectrum, as constrained by the COBE/Firas bounds on $\mu$ and $y$-distortions \citep{2013MNRAS.434.1619C}.}
\label{Fig1}
\end{figure}

In Fig.\,\ref{Fig1}, we have presented the modified matter power spectra derived from Eq.\,\ref{Eq1} for different values of $k_c$ and $n$, in comparison with the standard $\Lambda$CDM power spectrum. The data points represent observational constraints from UV luminosity function \citep{2022ApJ...928L..20S}, and Lyman-$\alpha$ forest measurements \citep{2019MNRAS.489.2247C}, while the olive curve marks the maximal cutoff scale for $k^4$ growth in the adiabatic curvature power spectrum, constrained by COBE/Firas bounds on $\mu$ and $y$-distortions \citep{2013MNRAS.434.1619C}.

As $k_c$ increases, the suppression of small-scale power becomes more confined to higher wavenumbers, which may reduce its impact on larger structures while remaining consistent with Lyman-$\alpha$ and UV luminosity function data. A higher spectral index $n$ can enhance the small-scale power, potentially increasing the likelihood of early halo formation \citep{2023PhRvD.107d3502H}. These adjustments may influence the abundance and distribution of dark matter halos, particularly at high masses, which can help account for the unexpectedly large number of massive galaxies observed by JWST at high redshifts. Hence, introducing suitable values for $k_c$ and $n$ may provide a viable means of reconciling theoretical predictions with recent high-redshift galaxy observations. Several well-motivated theoretical scenarios can generate the power spectrum modifications described by Eq.\,\ref{Eq1}:

\textit{Axion dark matter with post-inflationary symmetry breaking}: In models where axion-like particles constitute dark matter and the associated $U(1)$ Peccei-Quinn symmetry is broken after cosmic inflation, the Kibble mechanism generates large-amplitude fluctuations on small scales \citep{1993PhRvL..71.3051K}. This process leads to axion minicluster formation, contributing an $n = 0$ component to the matter power spectrum with amplitude \citep{2022PhRvD.106j3514E, 2022PhRvD.105e5025O}: 
\begin{equation} 
A = \frac{24 \pi^2}{5 D^2(z_{\text{eq}}) k_{\text{cut}}^{3}},
\end{equation}
where $D(z_{\text{eq}})$ is the growth factor at matter-radiation equality. The cutoff scale corresponds to the comoving wavenumber that entered the horizon when the axion field of mass $m_a$ began coherent oscillations: 
\begin{equation} 
k_{\text{cut}} \approx 300\,\text{Mpc}^{-1} \sqrt{\frac{m_a}{10^{-18}\,\text{eV}}}.
\end{equation}
By employing a power-law approximation for the $\mathcal{P}_{\Lambda\rm CDM}(k)$ matter power spectrum near $ k_c = \mathcal{O}(10)\, h\,\text{Mpc}^{-1} $, one can find that the characteristic scale $k_c$, at which $P_{\Lambda\text{CDM}}(k_c) = A$, is approximately given by:
\begin{equation}\label{kcone}
k_c \simeq 3\, h\,\text{Mpc}^{-1} \left( \frac{m_a}{10^{-18}\,\text{eV}} \right)^{0.6}
\end{equation}
This relation demonstrates that the condition $ k_{\text{cut}} \gtrsim k_c $ holds within the corresponding range of $k_c$.

\textit{PBH clustering}: Scenarios involving massive PBHs or their spatial clusters can generate shot noise contributions to the matter power spectrum \citep{2018MNRAS.478.3756C}. The discrete, Poisson-distributed nature of these compact objects produces an $n = 0$ enhancement with amplitude \citep{2019PhRvD.100h3528I, 2020JCAP...11..028D}: 
\begin{equation} 
A = 6\pi^2 \frac{f_{\text{PBH}}^2}{D^2(z_{\text{eq}}) k_{\text{cut}}^{3}},
\end{equation}
where $f_{\text{PBH}}$ represents the dark matter fraction in PBHs. The cutoff scale is determined by the mean separation between these objects: 
\begin{equation} 
k_{\text{cut}} \approx 900 \,h\,\text{Mpc}^{-1} \left(\frac{f_{\text{PBH}} 10^4\,M_{\odot}}{m_{\text{PBH}}}\right)^{1/3}.
\end{equation}
While individual massive PBHs face stringent constraints from CMB observations, scenarios involving clustered lighter BHs may circumvent these limitations while still producing the required power spectrum enhancement.

At spatial scales smaller than this cutoff, one anticipates the presence of only a single PBH within the comoving sphere. In this regime, the so-called seed effect becomes dominant over the Poissonian contribution \citep{1983ApJ...268....1C, 2022ApJ...926..205C}. The characteristic scale $k_c$ associated with this transition is approximately given by:
\begin{equation}
    k_c \simeq 6\, h\,\text{Mpc}^{-1} \left( \frac{f_{\text{PBH}} m_{\text{PBH}}}{10^4 M_\odot} \right)^{-0.4}
    \label{eq:kc_pbh}
\end{equation}
Similar to Eq.\,\ref{kcone}, this approximation performs well for $ k_c = \mathcal{O}(10)\, h\,\text{Mpc}^{-1} $. Thus, one can find that $k_c < k_{\text{cut}}$ holds when $f_{\text{PBH}} > 10^{-4} \left(m_{\text{PBH}}/10^4 M_\odot \right)^{-0.09}$.

However, constraints from CMB observations impose an upper bound of $f_{\text{PBH}}<10^{-8}$ for PBHs with masses around $10^4 M_\odot$ \citep{2020PhRvR...2b3204S, 2021RPPh...84k6902C}, indicating that explanations of JWST observations based solely on Poisson fluctuations from heavy PBHs are not viable under standard assumptions. That said, the accretion-related constraints may be relaxed if lighter PBHs formed in dense clusters, e.g., in \citep{2021PhRvL.126d1101F, 2021PhRvD.104l3539D, 2024PhRvD.110k5014J}, with individual cluster masses exceeding $10^4 M_\odot$. In such a scenario, the clusters themselves would act as sources of Poisson-like fluctuations rather than the individual PBHs \citep{2022PhRvL.129s1302D}.

Recent studies show that if PBHs, particularly supermassive PBHs with masses about $10^9 M_\odot$ with $f_{\rm PBH} \sim 10^{-3}$, are initially clustered beyond a Poisson distribution, the induced power spectrum acquires an additional scale-dependent enhancement \citep{2023arXiv230617577H}. The clustering contribution is characterized by a spatial correlation function $\xi_{\rm PBH}(x)$, leading to a modification:
\begin{equation}
P_\xi(k) = 4\pi\xi_0 \frac{\sin(kx_{\rm cl}) - kx_{\rm cl}\cos(kx_{\rm cl})}{k^3},
\end{equation}
where $x_{\rm cl}$ represents the comoving size of the gravitationally bound cluster\footnote{The spatial correlation function of PBHs can be modelled as \citep{2022PhRvL.129s1302D}
\begin{align}\label{eq:xix}
    \xi_{\rm PBH}(x)=
\begin{cases}
    \xi_0  \qquad {\rm if} \qquad x\le x_{\rm cl},  \\
    0 \qquad  \ \, {\rm otherwise},
\end{cases}
\end{align}
}. This relation dominates over the Poisson term on large scales ($k \ll x_{\rm cl}^{-1}$) and naturally shifts the effective enhancement to smaller $k$ values. In such cases, the relevant cutoff is governed by the cluster separation rather than individual PBH separation \citep{2023arXiv230617577H}:
\begin{equation}
k_{\text{cut, cluster}} \approx \left(\frac{3\pi}{2\xi_0}\right)^{1/3} \frac{1}{x_{\rm cl}},
\end{equation}
allowing structure growth to be significantly boosted at earlier times. 

Therefore, clustered PBH scenarios modify the matter power spectrum in a way that mimics Eq.~\eqref{Eq1}, but with physically motivated and extended support for $n \simeq 0$ enhancements over broader comoving scales. This framework offers a compelling explanation for observed excesses in early galaxy formation, particularly those reported by JWST, and relaxes the otherwise tight constraints on $f_{\text{PBH}}$ for massive objects by distributing their effect through clustering rather than isolated Poisson noise.

\textit{Inflationary spectral features}: Enhanced adiabatic power spectra can arise in single-field inflation models featuring potential structures that temporarily decelerate the inflaton field \citep{2017PDU....18...47G, 2017JCAP...09..020K, 2017PDU....18....6G}. Such scenarios typically generate curvature power spectra growing as $k^{5-|2-n_s|}$ above the characteristic scale, corresponding to $n \approx 0.14\mbox{-}0.68$ in the parametrization for $k_c$ ranging from $100\,h\,\text{Mpc}^{-1}$ to $1\,h\,\text{Mpc}^{-1}$, respectively \citep{2023JCAP...03..013K}. 

While a $k^{4}$ scaling of the curvature power spectrum is a characteristic feature in a broad class of inflationary models \citep{2019JCAP...06..028B}, steeper spectral indices can emerge in more specialized scenarios involving finely tuned dynamics or non-standard inflationary potentials \citep{2021PhRvD.103b3535T}. For the purposes of this analysis, we restrict our consideration to power-law spectra with spectral index $n\leq 1$ , corresponding to a moderate enhancement of small-scale fluctuations.

The first two scenarios involve isocurvature perturbations, which are less constrained observationally than adiabatic modes. COBE/FIRAS bounds on spectral distortions primarily restrict adiabatic fluctuations, while isocurvature constraints remain comparatively weak \cite{2013MNRAS.434.1619C}. In particular, the small-scale cutoffs relevant to axion miniclusters and PBH clustering lie well below the FIRAS-sensitive regime. For adiabatic enhancements arising from inflationary features, COBE/FIRAS excludes PBH formation for $k_c \lesssim 10^2\, h\,\text{Mpc}^{-1}$, except when the associated growth of the curvature power spectrum is sufficiently mild,i.e., $n < 0$. Nonetheless, scenarios with rapid power growth peaking near $k_c \sim \mathcal{O}(10)\, h\,\text{Mpc}^{-1}$ and ending within four orders of magnitude of the CMB amplitude remain allowed. Thus, the modified matter power spectra discussed here offer a viable framework to explore early structure formation, such as the emergence of massive galaxies observed by JWST, while remaining compatible with large-scale observational bounds.
\section{Halo Mass Functions}\label{sec:iii}
Analytical expressions for the halo mass function can be rigorously derived through excursion set theory, which treats the density field as a stochastic process operating across multiple scales. The theoretical underpinnings rely on the spherical collapse model, which establishes the critical overdensity threshold required for the gravitational collapse of spherically symmetric perturbations \citep{1974ApJ...187..425P}. For an Einstein-de Sitter Universe under the assumption of perfect spherical symmetry, this critical overdensity threshold is analytically expressed as:
\begin{equation}
\delta_{\rm sc}=\frac{3(12\pi)^{2/3}}{20}\left(1-0.01231\log\left[1+\frac{\Omega_{\rm m}^{-1}-1}{(1+z)^{3}}\right]\right),
\end{equation}
where $\Omega_{\rm m}$ denotes the matter density parameter. This threshold overdensity is typically approximated as $1.686$ over limited redshift intervals. However, a notable limitation of this conventional approach is that $\delta_{\rm sc}$ exhibits minimal mass dependence, potentially leading to systematic underestimations in certain mass regimes.

\begin{figure*}
\centering
\includegraphics[width=1\linewidth]{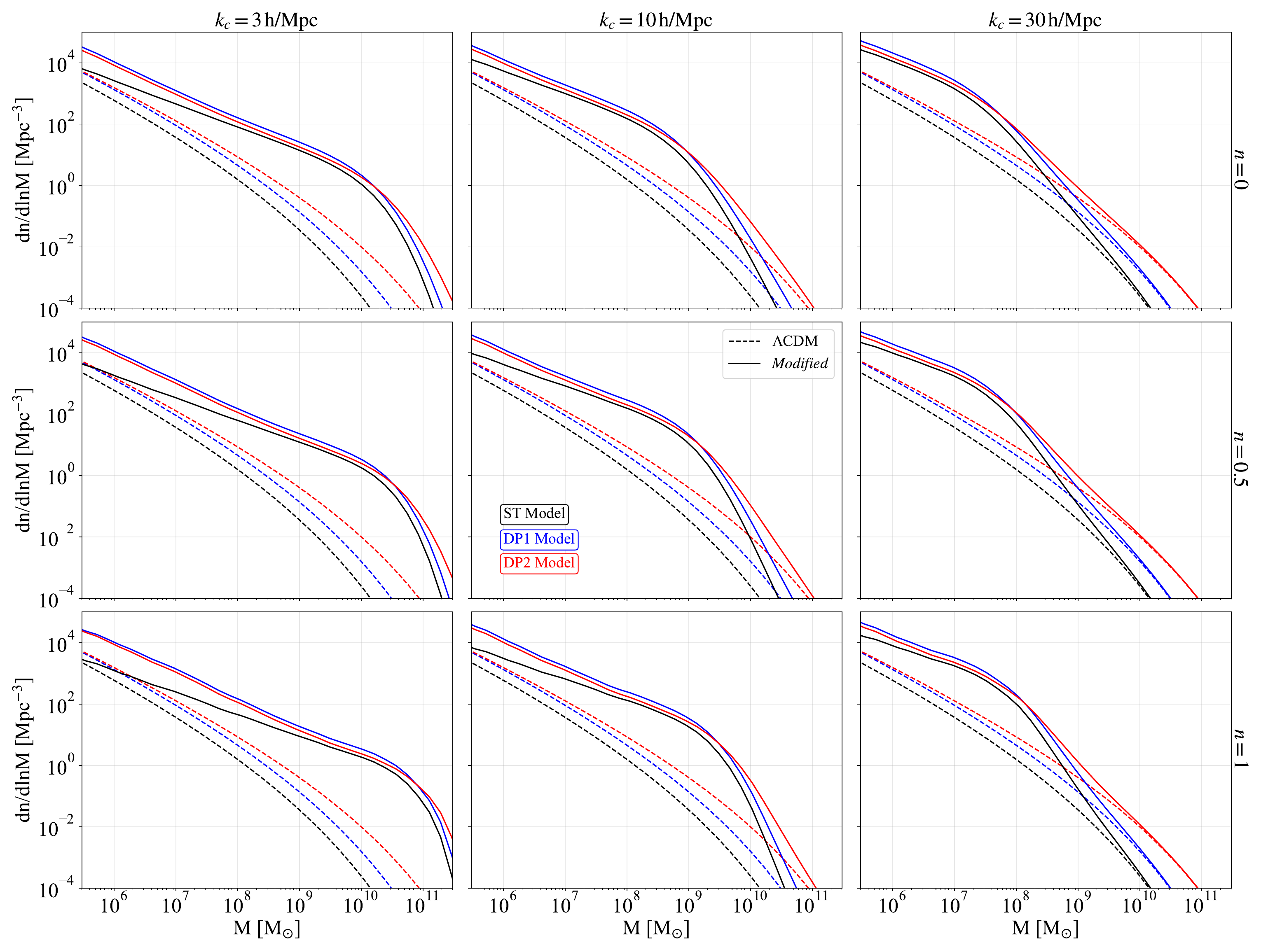}
\caption{Comparison of the ST, DP1, and DP2 halo mass functions, computed using the modified power spectrum of Eq.,(\ref{Eq1}), with those from the standard $\Lambda$CDM power spectrum, shown as a function of mass $M$. The comparison is presented for different values of the characteristic scale, $k_c = 3$, $10$, and $30\,h\,{\rm Mpc}^{-1}$, and spectral index, $n = 0$, 0.5, and 1, at a fixed redshift of $z = 10$.}
\label{Fig2}
\end{figure*}

The first significant refinement to the PS threshold was pioneered in \citep{1998A&A...337...96D}, demonstrating that the collapse threshold can incorporate mass-dependent corrections of the form:
\begin{equation}
\delta_{\rm cm}=\delta_{\rm sc}\left(1+\frac{\beta}{\nu^{\alpha}}\right),
\end{equation}
with best-fit parameter values $\alpha=0.585$ and $\beta=0.46$. Subsequently, \citep{2001MNRAS.323....1S} derived an alternative threshold by considering ellipsoidal rather than spherical collapse geometry:
\begin{equation}
\delta_{\rm ec}=\delta_{\rm sc}\left(1+\frac{\beta_1}{\nu^{\alpha_1}}\right),
\end{equation}
where $\alpha_1=0.615$, $\beta_1=0.485$, and $\nu$ represents the peak height, rigorously defined as:
\begin{equation}
\nu \equiv \frac{\delta_{\rm sc}(z)}{\sigma_M}=\frac{1.686}{D(z)\sigma_M}.
\end{equation}
In the above relation, $\sigma_M$ represents the linear root-mean-square density fluctuation on comoving scales encompassing halo mass $M$:
\begin{equation} 
\sigma^2_{M} = \frac{1}{2\pi^{2}}\int_{0}^{\infty} \mathcal{P}_{\rm Mod}(k) W^2(kR) k^{2} {\rm d}k,
\end{equation}
where $W(kR)$ denotes the filtering window function smoothed over the comoving scale $R$. Note that the halo mass can be related to comoving scale as $M=4\pi \bar{\rho}_{\rm m,0} R^3/3$, in which $\bar{\rho}_{\rm m,0}$ is the present-day comoving mean density of dark matter. In this work, we have employed a real-space top-hat window function defined as $W(x)=3(\sin x - x\cos x)/x^3$.

The excursion set formalism establishes the foundation for calculating the unconditional mass function of dark matter halos, which quantifies the mean comoving number density of halos per logarithmic mass bin. This fundamental relationship is expressed as:
\begin{equation} 
n(M, z) = \frac{\bar{\rho}_{\rm m,0}}{M^2} \left|\frac{{\rm d}\log\nu}{{\rm d}\log M} \right| \nu f(\nu),
\end{equation}
where $\nu f(\nu)$ constitutes the multiplicity function, characterizing the first-crossing distribution. The original Press-Schechter formalism yields the following analytical expression for the multiplicity function \citep{1974ApJ...187..425P}:
\begin{equation}
[\nu f(\nu)]_{\rm PS}=\sqrt{\frac{2}{\pi}}\frac{(\nu+0.556)\exp[-0.5(1+\nu^{1.34})^{2}]}{(1+0.0225\nu^{-2})^{0.15}}.
\end{equation}

Nevertheless, substantial discrepancies persist between PS theoretical predictions and observed dark matter halo distributions from high-resolution numerical simulations. These discrepancies arise from various physical processes not adequately accounted for in the original PS formalism, yet these processes substantially influence halo abundance statistics. The ST mass function addresses these limitations by incorporating geometric corrections and extending the spherical collapse framework to more realistic ellipsoidal collapse models. The ST formulation takes the analytical form \citep{2001MNRAS.323....1S}:
\begin{equation}
[\nu f(\nu)]_{\rm ST}=A_{1}\sqrt{\frac{2\nu^{\prime}}{\pi}}\left(1+\frac{1}{\nu^{\prime q}}\right)\exp\left(-\frac{\nu^{\prime}}{2}\right),
\end{equation}
where $q=0.3$, $\nu^\prime=0.707\nu^2$, and $A_{1}=0.322$, with the normalization constant determined by requiring $\int f(\nu) d\nu = 1$.

Beyond geometric considerations during halo virialization processes, additional physical mechanisms significantly affect overdensity collapse dynamics and, consequently, the halo mass function. Incorporating these mechanisms is essential as they represent the fundamental physics of halo formation and evolution, as well as the underlying processes driving hierarchical structure formation throughout cosmic history. This approach enables the collapse threshold to depend on physically motivated parameters, allowing the barrier to adapt dynamically and yielding more accurate theoretical predictions for halo collapse. Critical corrections include the effects of angular momentum, dynamical friction processes, and cosmological constant \footnote{Although the cosmological constant contributes negligibly at high redshifts, it is retained for the sake of theoretical completeness.}. These refinements help minimize systematic discrepancies, particularly in contested mass ranges where observational constraints remain challenging.

When rigorously accounting for angular momentum and cosmological constant effects, the resulting mass function, designated DP1 in this investigation, provides enhanced theoretical accuracy \citep{2006ApJ...637...12D}:
\begin{equation}
[\nu f(\nu)]_{\rm DP1}=A_{2}\sqrt{\frac{\nu^{\prime}}{2\pi}}k(\nu^{\prime})\exp\left[-0.4019\nu^{\prime} l(\nu^{\prime})\right],
\end{equation}
where $A_{2}=0.974$ represents the normalization constant, and:
\begin{equation}
k(\nu^\prime)=\left(1+\frac{0.1218}{\nu^{\prime 0.585}}+\frac{0.0079}{\nu^{\prime 0.4}}\right),
\end{equation}
\begin{equation}
l(\nu^\prime)=\left(1+\frac{0.5526}{\nu^{\prime 0.585}}+\frac{0.02}{\nu^{\prime 0.4}}\right)^2.
\end{equation}

The influence of dynamical friction on the collapse barrier was systematically investigated in \citep{2017JCAP...03..032D}, yielding the mass function denoted DP2:
\begin{equation}
[\nu f(\nu)]_{\rm DP2}=A_{3}\sqrt{\frac{\nu^\prime}{2\pi}}m(\nu^\prime)\exp[-0.305\nu^{\prime2.12} n(\nu^\prime)],
\end{equation}
where $A_{3}=0.937$ represents the normalization factor, and:
\begin{equation}
m(\nu^\prime)=\left(1+\frac{0.1218}{\nu^{\prime 0.585}}+\frac{0.0079}{\nu^{\prime 0.4}}+\frac{0.1}{\nu^{\prime 0.45}}\right),
\end{equation}
\begin{equation}\label{nnuprime}
n(\nu^\prime)=\left(1+\frac{0.5526}{\nu^{\prime 0.585}}+\frac{0.02}{\nu^{\prime 0.4}}+\frac{0.07}{\nu^{\prime 0.45}}\right)^2.
\end{equation}

Modifications to the matter power spectrum directly influence the mass variance $\sigma_M$ and consequently alter the predicted halo mass function through the peak height parameter $\nu$. Suppression of small-scale power, as explored in Eq.~(\ref{Eq1}), systematically reduces $\sigma_M$ for low-mass halos, increasing their peak height $\nu$ and exponentially suppressing their abundance while enhancing the relative contribution of high-mass halos. These effects are amplified at high redshifts through the linear growth factor $D(z)$, where the formation of massive halos becomes critically sensitive to power spectrum modifications since these rare objects occupy the exponential tail of the mass function where small changes in $\sigma_M$ produce order-of-magnitude variations in abundance, a sensitivity particularly important for explaining the observed properties of high-redshift galaxies that require efficient assembly of massive dark matter halos \cite{2023ApJS..265....5H}.

In Fig.\,\ref{Fig2}, we have analyzed the impact of a modified matter power spectrum on the halo mass function using three different formalisms: ST, DP1, and DP2. We have considered a range of characteristic scales $k_c = 3, 10$, and $30\,h\,\text{Mpc}^{-1}$, and spectral indices $n=0,0.5$, and $1$, all evaluated at redshift $z = 10$. We have observed that, across all models, introducing small-scale power enhancements through increasing $n$ or decreasing $k_c$ has led to a suppression of low-mass halos and a simultaneous boost in the abundance of high-mass halos. This effect is particularly prominent in the DP2 model, which incorporates additional collapse physics such as dynamical friction. Compared to the ST model, both DP1 and DP2 have resulted in more massive halos, especially for $n = 1$, indicating that enhanced power on small scales may significantly influence early structure formation.

We have also compared the response of the three halo mass function models to changes in $k_c$ and $n$. We have found that the DP2 model consistently exhibits the strongest amplification in the high-mass tail, followed by DP1, with ST showing the least sensitivity to the modified spectrum. For lower values of $k_c$, where the modification affects broader comoving scales, we have seen stronger deviations from the standard $\Lambda$CDM scenario, particularly in the high-mass tail. In contrast, higher $k_c$ values have localized the modifications to smaller scales, leading to more moderate changes in the halo mass function. Notably, as $n$8 increases, the effect of enhanced small-scale clustering becomes more pronounced in the high-mass tail, particularly in the DP1 and DP2 models.

\section{Stellar Mass Density and JWST Observations}\label{sec:iv}
Once the halo mass function is determined, a closely associated quantity of interest is the cumulative halo mass density, denoted as $\rho_m(> M, z)$, which quantifies the total mass contained in halos above a given mass threshold at redshift $z$. It is defined as:
\begin{equation}
\rho_m(> M, z) = \int_{M}^{\infty}M^{\prime}\frac{{\rm d}n(M^{\prime}, z)}{{\rm d}M^{\prime}}{\rm d}M^{\prime}.
\end{equation}
Under the assumption that the maximum stellar content of a halo is constrained by its cosmological allotment of baryons, an upper limit on the stellar mass can be expressed as $M_{\star, \mathrm{max}} = f_\star f_b M$, where $f_b=0.156$ denotes the cosmic baryon fraction \footnote{The osmic baryon fraction is defined as $f_b\equiv \Omega_b / \Omega_m$, where $\Omega_b$ and $\Omega_m$ correspond to the baryonic and total matter density parameters, respectively.} \citep{2020A&A...641A...6P}, and $f_\star \leq 1$ represents the star formation efficiency.

For the purpose of interpreting JWST observations, a key quantity is the cumulative stellar mass density in galaxies with stellar mass greater than $M_\star$, denoted by $\rho_\star(> M_\star, z)$, and defined as \citep{2023NatAs...7..731B}:
\begin{equation}
\rho(> M_\star, z)=f_\star f_b \int_{M_\star/(\epsilon f_b)}^{\infty} M \frac{{\rm d}n(M, z)}{{\rm d}M} {\rm d}M.
\end{equation}
At a given redshift, the cumulative stellar mass density in galaxies with stellar mass greater than $M_\star$ is constrained by the upper bound $\rho_\star(> M_\star) \leq f_\star  f_b\, \rho_m(> M_\star / f_b)$, reflecting the maximal conversion of baryonic matter into stars within halos.

\subsection{Case I: $z\simeq 10$}
\begin{figure*}
\centering
\includegraphics[width=1\linewidth]{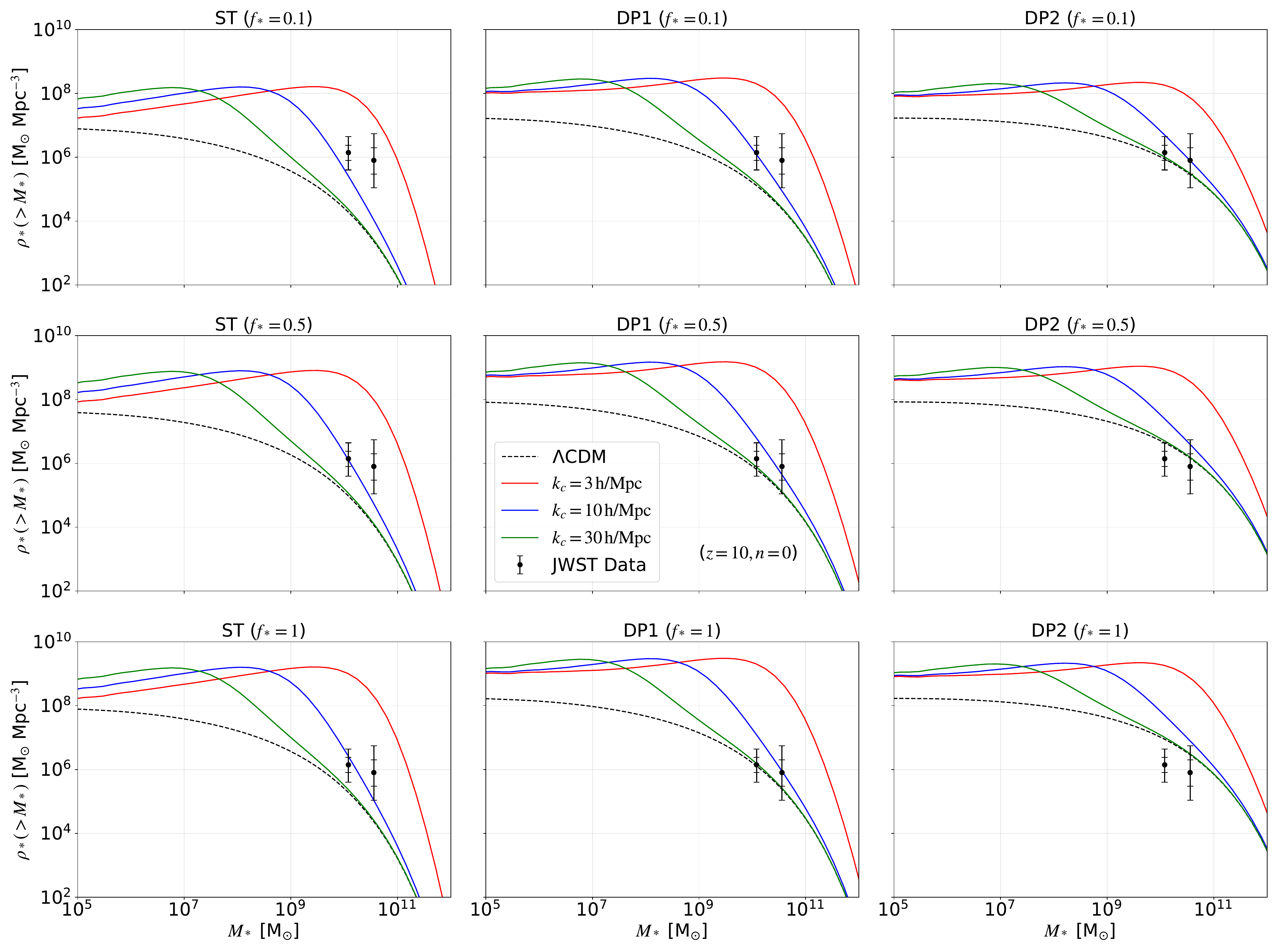}
\caption{The cumulative stellar mass density for galaxies with stellar masses exceeding a mass $M_\star$ at redshift $z = 10$ is computed based on three distinct halo mass functions: ST, DP1, and DP2. These calculations incorporate a modified power spectrum as indicated in Eq.\,(\ref{Eq1}) with spectral index $n = 0$, in comparison with the one obtained from the standard $\Lambda$CDM power spectrum. The analysis is conducted for various characteristic scales, specifically $k_c = 3, 10,$ and $30\, h\, {\rm Mpc}^{-1}$, and assumes a range of star formation efficiencies $f_\star = 0.1, 0.5$, and $1.0$. For comparison, observational constraints derived from JWST data \citep{2023Natur.616..266L} at $z \simeq 10$ are included, with the black error bars representing the $1\sigma$ and $2\sigma$ confidence levels of the inferred stellar mass density.}
\label{Fig3a}
\end{figure*}

In Fig.\,\ref{Fig3a}, we have illustrated the cumulative stellar mass density $\rho_\star(>M_\star)$ at $z = 10$, computed using three distinct halo mass function models: ST, DP1, and DP2. Each model was evaluated under both the standard $\Lambda$CDM power spectrum and modified power spectrum with characteristic scales $k_c = 3,\,10,$ and $30\,h\,\mathrm{Mpc}^{-1}$, assuming a fixed spectral index $n=0$. We have alos shown these predictions for various star formation efficiencies $f_\star = 0.1,\ 0.5,\ 1.0$. Moreover, we have included two cumulative stellar mass densities derived from the JWST observations for $M_\star \gtrsim 10^{10} M_\odot$ and at $z\simeq 10$ that are $\rho_\star = 1.3^{+1.1}_{-0.6} \times 10^6\,M_\odot\,\mathrm{Mpc}^{-3}$, and $9^{+11}_{-6} \times 10^5\,M_\odot\,\mathrm{Mpc}^{-3}$. These are represented in the plots as black dots with 1$\sigma$ and 2$\sigma$ error bars and allow a direct comparison between theoretical predictions and high-redshift galaxy observations.

In the ST model, one can observe that predictions based on the standard $\Lambda$CDM power spectrum consistently underpredict the JWST-inferred stellar mass density across all values of $f_\star$. Incorporation of the modified power spectrum leads to an increase in the cumulative stellar mass density, particularly for lower values of $k_c$. At almost all values of $f_\star$, the $3\,h\,\mathrm{Mpc}^{-1}\leq k_c\leq 10\,h\,\mathrm{Mpc}^{-1}$ case approaches the lower 2$\sigma$ bound of JWST data. This highlights the limited capability of the ST mass function within the standard $\Lambda$CDM model, which, even under optimal conditions, struggles to explain the rapid early stellar buildup observed by JWST. However, theoretical predictions derived from the ST formalism, when applied to scenarios incorporating modifications to the power spectrum, may exhibit compatibility with observational data obtained by JWST at cosmological redshifts of approximately $z \simeq 10$.

In contrast, the DP1 model, which accounts for additional physical parameters such as angular momentum, exhibits significantly enhanced predictions for the cumulative stellar mass density. Under the assumption of a standard $\Lambda$CDM power spectrum, the DP1 formalism produces theoretical predictions that remain statistically consistent with stellar mass density measurements from JWST, exhibiting discrepancies of less than $2\sigma$ (with some cases showing less than $1\sigma$ tension) for $0.5\leq f_\star \leq 1$. Specifically, the utilization of more realistic DP1 mass function that encapsulates key physical parameters may facilitate better agreement between theoretical predictions of cumulative stellar mass density within the standard $\Lambda$CDM model and JWST observational data. 

Furthermore, the cumulative stellar mass density obtained through the DP1 mass function within the modified power spectrum framework yields compelling findings. In this context, as illustrated in the figure, theoretical predictions for the stellar mass density with $f_\star=0.1$ and $3\,h\,\mathrm{Mpc}^{-1}\leq k_c\leq 10\,h\,\mathrm{Mpc}^{-1}$ exhibit potential consistency with the observational data from the JWST. Furthermore, when considering the parameter ranges $0.5\leq f_\star\leq 1$ and $10\,h\,\mathrm{Mpc}^{-1}\leq k_c\leq 30\,h\,\mathrm{Mpc}^{-1}$ , a notable agreement (less than $1\sigma$ tension) emerges between the predictions derived from the DP1 halo mass function and the JWST observations. This indicates that the DP1 mass function is broadly consistent with JWST data in regimes characterized by lower star formation efficiencies and relatively larger-scale modes. However, achieving consistency at smaller physical scales, corresponding to higher $k_c$, requires the inclusion of higher star formation efficiencies, suggesting that enhanced star formation activity is necessary to reconcile theoretical models with observations in these regimes.

\begin{figure*}
\centering
\includegraphics[width=1\linewidth]{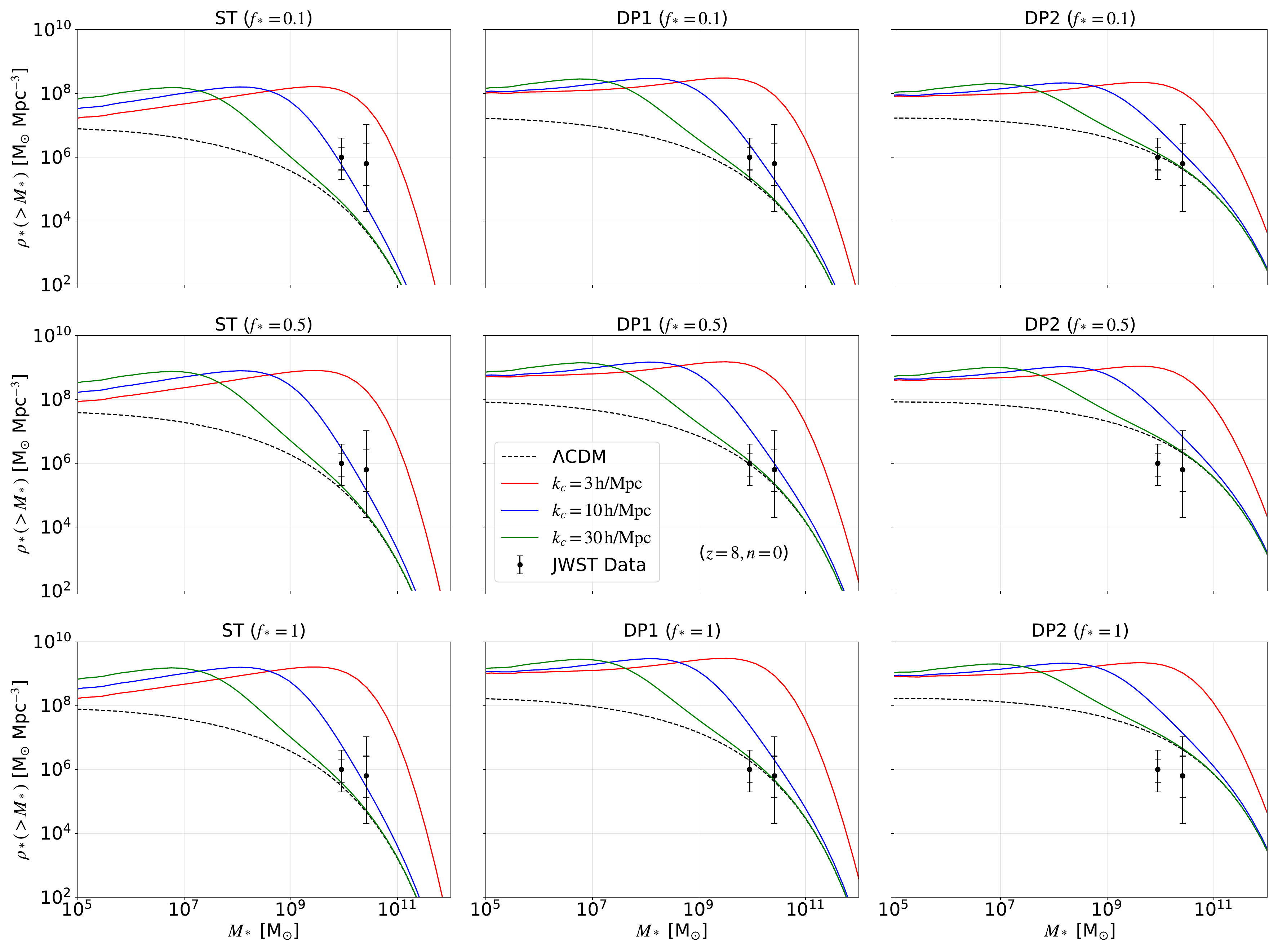}
\caption{Similar to Fig.\,\ref{Fig3a} but at redshift $z=8$. Also, observational constraints derived from JWST data \citep{2023Natur.616..266L} at $z \simeq 8$ are depicted with the black error bars representing the $1\sigma$ and $2\sigma$ confidence levels.}
\label{Fig4a}
\end{figure*}

We have also presented the theoretical predictions derived from the DP2 mass function for cumulative stellar mass density in the third column of the figure, which incorporates the effects of both angular momentum and dynamical friction parameters. It is evident that the DP2 mass function, when applied to scenarios characterized by star formation efficiencies $0.1\leq f_\star\leq 0.5$ (corresponding to moderate star formation efficiency) within the framework of the standard $\Lambda$CDM model, exhibits suitable concordance with observational constraints derived from JWST. Specifically, one can see that the model predictions maintain consistency with JWST observations, with deviations typically confined within $1\sigma$ confidence levels and, in certain instances, remaining below $2\sigma$ confidence levels. Meanwhile, as the star formation efficiency approaches unity, the DP2 mass function predicts stellar accretion mass densities that increasingly deviate from the constraints derived from JWST observations. While these predictions may still be consistent in a certain case, they exhibit deviations exceeding $2\sigma$ in other case.

Moreover, the DP2 mass function yields compelling theoretical predictions for the cumulative stellar mass density when implemented within a modified power spectrum framework. Through alterations to the power spectrum at intermediate scales ($10\,h\,\mathrm{Mpc}^{-1}\leq k_c\leq 30\,h\,\mathrm{Mpc}^{-1}$) and under the assumption of lower star formation efficiency ($f_\star = 0.1$), this formulation achieves remarkable concordance with JWST observational data at $z \simeq 10$, maintaining statistical consistency within $1\sigma$ confidence levels. Nevertheless, a systematic deviation in model-data agreement is observed as star formation efficiency parameters increase as $f_\star \rightarrow 1$.

We have further extended our analysis in Appendix by examining two additional scenarios corresponding to spectral indices $n=0.5$ and $n=1$ at redshift $z \simeq 10$, as shown in Figs.\,\ref{Fig3b} and \ref{Fig3c}, respectively. These supplementary cases clearly illustrate that increasing the spectral index systematically amplifies the predicted cumulative stellar mass density across all three halo mass function models, ST, DP1, and DP2, within the modified power spectrum framework. Compared to the baseline $n=0$ case, the enhancement is particularly significant for massive galaxies and becomes more pronounced at larger characteristic scales $k_c$, consistent with the theoretical expectation that steeper small-scale spectral slopes boost high-mass halo formation. Notably, for $n=0.5$, the resulting cumulative stellar mass densities exhibit improved agreement with JWST observations, especially in the DP1 and DP2 models, while the case of $n=1$ shows a further amplification with only marginal deviations from the $n=0.5$ results. These findings underscore the exponential sensitivity of early massive galaxy formation to the small-scale behavior of the  power spectrum (see Appendix for detailed discussion).

\subsection{Case II: $z\simeq 8$}

In Fig.\,\ref{Fig4a}, we have presented the cumulative stellar mass density at redshift $z = 8$, computed using three distinct halo mass function models: ST, DP1, and DP2. Similar to the previous section, each of these models is examined under both the standard $\Lambda$CDM power spectrum and a modified power spectrum incorporating characteristic scales at $k_c = 3$, $10$, and $30\,h\,\mathrm{Mpc}^{-1}$, with the spectral index fixed at $n = 0$. The corresponding theoretical predictions are shown for a range of star formation efficiencies, specifically $f_\star = 0.1$, $0.5$, and $1.0$. To assess the viability of these models, we include observational estimates of the cumulative stellar mass density from the JWST for galaxies with $M_\star \gtrsim 10^{10}\,M_\odot$ at $z \simeq 8$. These estimates are given as $\rho_\star = 1.0^{+1.1}_{-0.5} \times 10^6\,M_\odot\,\mathrm{Mpc}^{-3}$ and $6.7^{+22}_{-5.1} \times 10^5\,M_\odot\,\mathrm{Mpc}^{-3}$. As in Fig.\,\ref{Fig3a}, the JWST observations of galaxy candidates at $z \simeq 8$ are depicted in this figure as black markers, accompanied by the $1\sigma$ and $2\sigma$ uncertainty intervals.

In the framework of the $\Lambda$CDM paradigm, the ST model demonstrates a systematic propensity toward underpredicting the empirically determined stellar mass density and achieves concordance with the high-redshift galaxy population detected by JWST exclusively under scenarios invoking elevated star formation efficiencies (specifically, $f_\star \rightarrow 1$), albeit with persistent discrepancies at the sub-$2\sigma$ statistical threshold. The implementation of augmented small-scale fluctuations via spectral modifications presents a viable mechanism for mitigating these observational tensions. In particular, such spectral alterations enable the ST framework to approach the $2\sigma$ confidence interval for $f_\star \geq 0.1$ and $3\,h\,\mathrm{Mpc}^{-1}< k_c\leq 10\,h\,\mathrm{Mpc}^{-1}$, while approximating the lower boundary of the $1\sigma$ observational constraint when $0.5 \leq f_\star \leq 1$ and $10\,h\,\mathrm{Mpc}^{-1}\leq k_c\leq 30\,h\,\mathrm{Mpc}^{-1}$. These findings indicate marginal progress in model-observation reconciliation when incorporating elevated star formation efficiencies alongside small-scale power spectrum enhancements. Nevertheless, even at $z \simeq 8$, the ST formalism exhibits fundamental limitations in reproducing the requisite stellar mass density within the standard $\Lambda$CDM theoretical construct.

The DP1 mass function exhibits enhanced predictive capability relative to the ST formalism through its incorporation of angular momentum considerations. Within the framework of the standard $\Lambda$CDM model, the cumulative stellar mass density computed via the DP1 prescription systematically exceeds that obtained from ST calculations and demonstrates remarkable concordance with JWST observational constraints, particularly for moderate to high star formation efficiencies ($0.5 \leq f_\star \leq 1$). This concordance indicates that physically motivated refinements to halo mass function parameterizations may yield substantially improved theoretical predictions within canonical $\Lambda$CDM cosmology. Furthermore, when implemented in conjunction with the modified power spectrum formulation, DP1 model outputs exhibit additional enhancement and demonstrate superior agreement with JWST data. Specifically, at $f_\star=0.1$, theoretical predictions incorporating $k_c \sim 10 \,h\,\mathrm{Mpc}^{-1}$ fall within $1\sigma$ observational uncertainties. Analogously, the DP1 mass function evaluated at $0.5\leq f_\star\leq 1$ with corresponding $10\,h\,\mathrm{Mpc}^{-1}\leq k_c\leq 30\,h\,\mathrm{Mpc}^{-1}$ values yields predictions that remain comfortably consistent with observational bounds. These results suggest that at $z \simeq 8$, the DP1 formalism is able to provide a viable theoretical framework for reconciling observed high-redshift stellar formation processes.

We have also displayed the theoretical predictions obtained from the DP2 mass function for cumulative stellar mass density in the third column of the figure. It is clear that the DP2 mass function, when implemented in scenarios characterized by star formation efficiencies $f_\star = 0.1$ within the framework of the standard $\Lambda$CDM model, shows excellent agreement with observational constraints obtained from JWST. Specifically, one can observe that the model predictions remain consistent with JWST observations, with deviations typically limited within $1\sigma$ confidence levels. However, as the star formation efficiency increases, the DP2 model shows systematic overprediction relative to observational constraints obtained from JWST data and demonstrates increased deviations from data. While these predictions may still be consistent in a certain case, they show deviations exceeding $2\sigma$ in other case.

Furthermore, the DP2 mass function produces compelling theoretical predictions for the cumulative stellar mass density when applied within a modified power spectrum framework. Through modifications to the power spectrum at intermediate scales ($10,h,\mathrm{Mpc}^{-1}\leq k_c\leq 30,h,\mathrm{Mpc}^{-1}$) and under the assumption of lower star formation efficiency ($f_\star = 0.1$), this formulation achieves remarkable agreement with JWST observational data at $z \simeq 8$, maintaining statistical consistency within $1\sigma$ confidence levels. Nevertheless, a systematic deviation in model-data agreement is observed as star formation efficiency parameters increase as $f_\star \rightarrow 1$.

In line with the preceding analysis, in Appendix, we have also presented two additional cases corresponding to spectral indices $n = 0.5$ and $n = 1$ at $z \simeq 8$, shown in Figs.\,\ref{Fig4b} and \ref{Fig4c}, respectively. These extended scenarios reveal that increasing the spectral index systematically enhances the cumulative stellar mass density across all three halo mass function models. The amplification is especially notable at higher characteristic scales $k_c$, where the predicted stellar mass densities exhibit improved agreement with JWST observations. These results further underscore the sensitivity of early structure formation to small-scale power spectrum modifications and reinforce the trend observed at $z \simeq 10$ (see Appendix for detailed discussion).

\section{Conclusions}\label{sec:v}
In this work, we have systematically investigated the tension between unexpectedly massive galaxy candidates observed by JWST at $z \gtrsim 8$ and standard $\Lambda\text{CDM}$ predictions by employing more physically realistic dark matter halo models combined with parametrically modified matter power spectra. We implemented three distinct halo mass function models, the conventional ST formalism and two physically motivated alternatives DP1 and DP2 that incorporate angular momentum, dynamical friction, and cosmological constant effects, alongside power spectrum modifications featuring small-scale enhancements characterized by spectral indices $0\leq n \leq 1$ and characteristic scales $3\,h\,\text{Mpc}^{-1} \leq k_c \leq 30\,h\,\text{Mpc}^{-1}$.

Our analysis of cumulative stellar mass densities at $z \simeq 8\text{--}10$ reveals several key findings that can improve our understanding of early galaxy formation. The conventional ST mass function systematically underpredicts JWST observations, achieving only marginal consistency within $2\sigma$ confidence levels for extreme star formation efficiencies $(f_\star \to 1)$ and specific power spectrum modifications. This demonstrates fundamental limitations in traditional halo formation models when applied to early cosmic epochs.

On the other hand, the DP1 and DP2 models, incorporating more complete collapse physics, exhibit improved agreement with observations. Remarkably, even within the standard $\Lambda\text{CDM}$ power spectrum, these models achieve statistical consistency within $1\text{--}2\sigma$ for moderate star formation efficiencies $(0.1 \leq f_\star \leq 0.5)$, suggesting that neglected physical processes in halo formation may partially resolve the JWST tension without requiring exotic modifications to cosmology.

When realistic halo models are coupled with modified power spectra, one can observe enhancements in predicted stellar mass densities. The DP1 model demonstrates consistency within $1\sigma$ observational constraints across broad parameter ranges, particularly for intermediate scales $(10\,h\,\text{Mpc}^{-1} \lesssim k_c \lesssim 30\,h\,\text{Mpc}^{-1})$ and moderate efficiencies. The DP2 model shows similar improvements but exhibits sensitivity to high star formation efficiencies, occasionally overpredicting observations for $(f_\star \to 1)$.

Our framework demonstrates that several well-motivated theoretical scenarios, including axion minicluster formation $(n = 0)$, clustered primordial black holes $(n = 0)$, and inflationary spectral features $(n \simeq 0.14\text{--}0.68)$, can generate the required small-scale power enhancements while remaining consistent with existing observational constraints. These modifications primarily affect the exponentially sensitive high-mass tail of the halo mass function, enabling efficient formation of massive dark matter halos at early times.

Moreover, our analysis reveals systematic dependencies on both characteristic scale and spectral index, with steeper spectral slopes (higher $n$) producing more pronounced amplifications in massive galaxy formation. The effectiveness of different halo models varies with scale: DP1 performs optimally at intermediate scales, while DP2 shows enhanced sensitivity to small-scale modifications.

Our findings suggest that incorporating realistic halo collapse physics, often neglected in standard analyses, represents a crucial yet underexplored avenue for reconciling JWST observations with $\Lambda\text{CDM}$ cosmology. This approach offers a more conservative alternative to exotic dark matter models or radical modifications to cosmological parameters. The success of DP1 and DP2 models in reproducing observations with moderate star formation efficiencies indicates that the apparent crisis in early galaxy formation may be partially attributable to oversimplified theoretical frameworks rather than fundamental failures of the standard cosmological model.

Although the results are encouraging, several uncertainties may still influence our interpretations, particularly concerning the calibration of star formation efficiency, the role of baryonic feedback, and the underlying mechanisms responsible for modifications to the matter power spectrum. The observed agreement between theory and JWST data could potentially arise from a degeneracy between enhanced halo formation and elevated star formation efficiency, making it difficult to disentangle the individual contributions without further constraints. Addressing these challenges will likely require a systematic calibration of improved halo models using high-resolution simulations with realistic baryonic physics, along with independent observational validation through multi-wavelength studies combining JWST, ALMA, and gravitational lensing. Furthermore, developing theoretical frameworks grounded in first principles for the origin of power spectrum modifications, and exploring their implications within alternative cosmologies, such as those involving modified gravity or non-standard dark energy, can help determine whether the JWST findings truly challenge the $\Lambda\text{CDM}$ paradigm or simply reflect the need to incorporate previously overlooked physical processes in early structure formation.
\begin{figure*}
\centering
\includegraphics[width=1\linewidth]{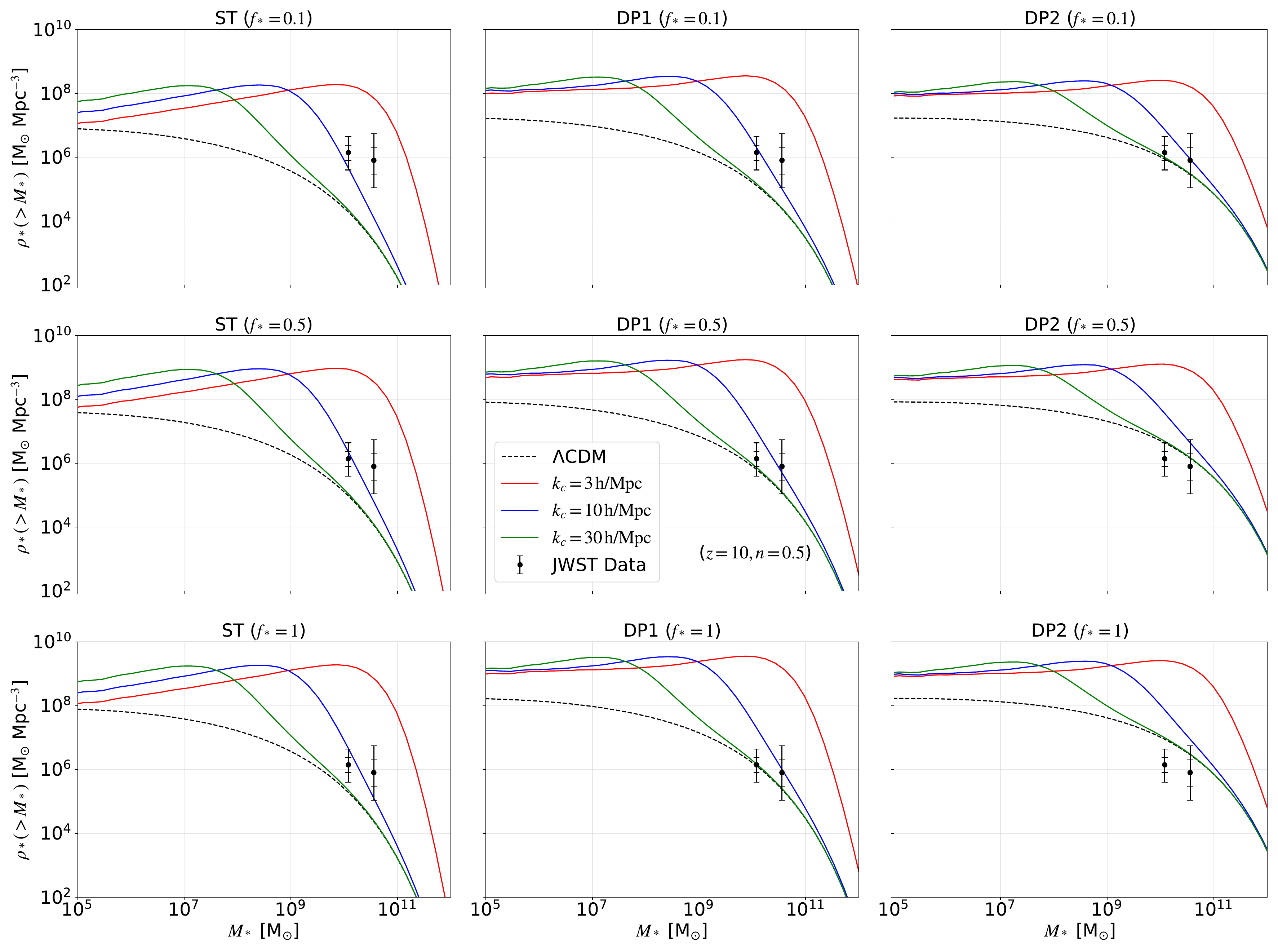}
\caption{Similar to Fig.\,\ref{Fig3a} but with $n=0.5$.}
\label{Fig3b}
\end{figure*}

\begin{figure*}
\centering
\includegraphics[width=1\linewidth]{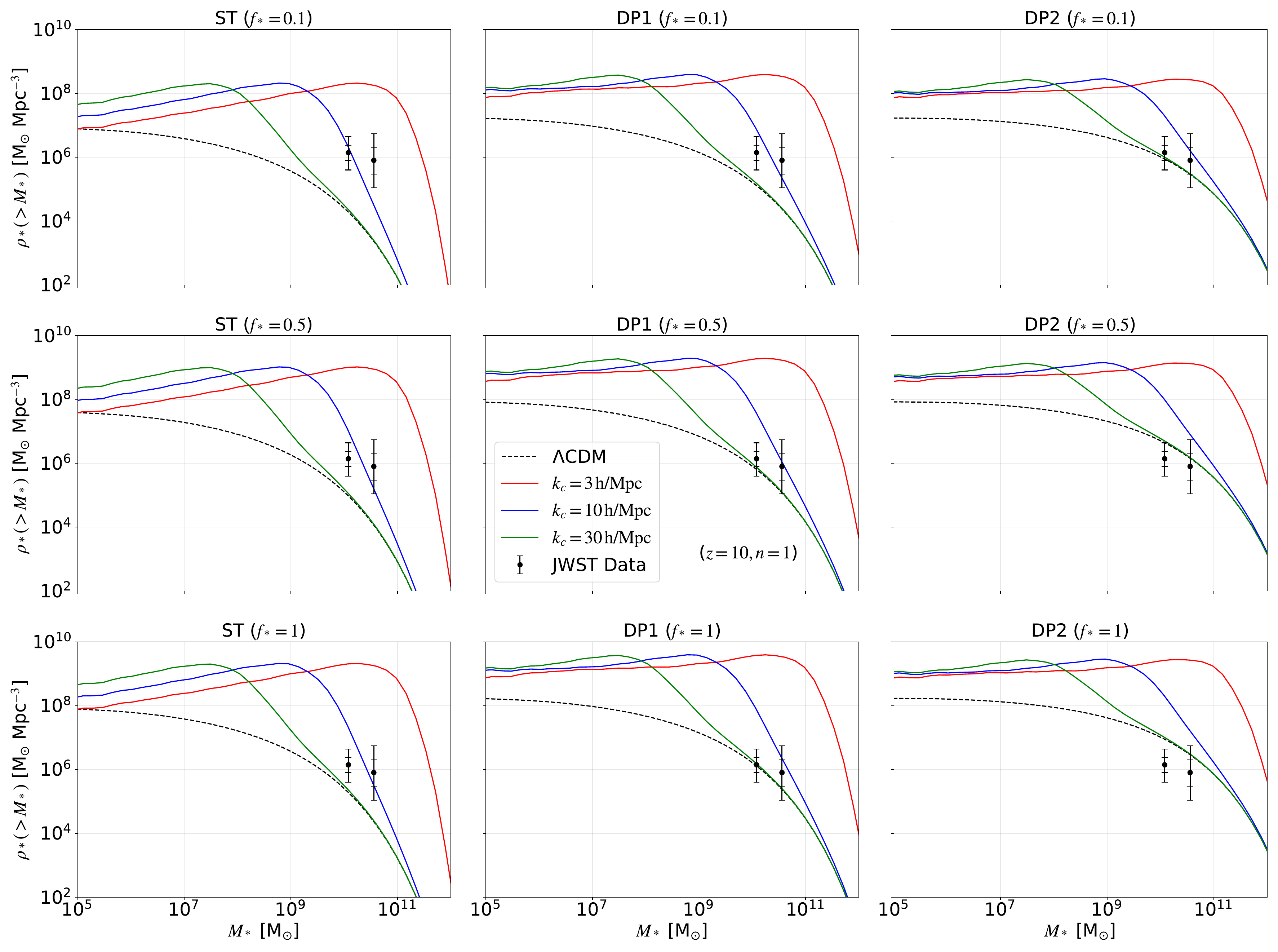} 
\caption{Similar to Fig.\,\ref{Fig3a} but with $n=1$.}
\label{Fig3c}
\end{figure*}

 \begin{figure*}
\centering
\includegraphics[width=1\linewidth]{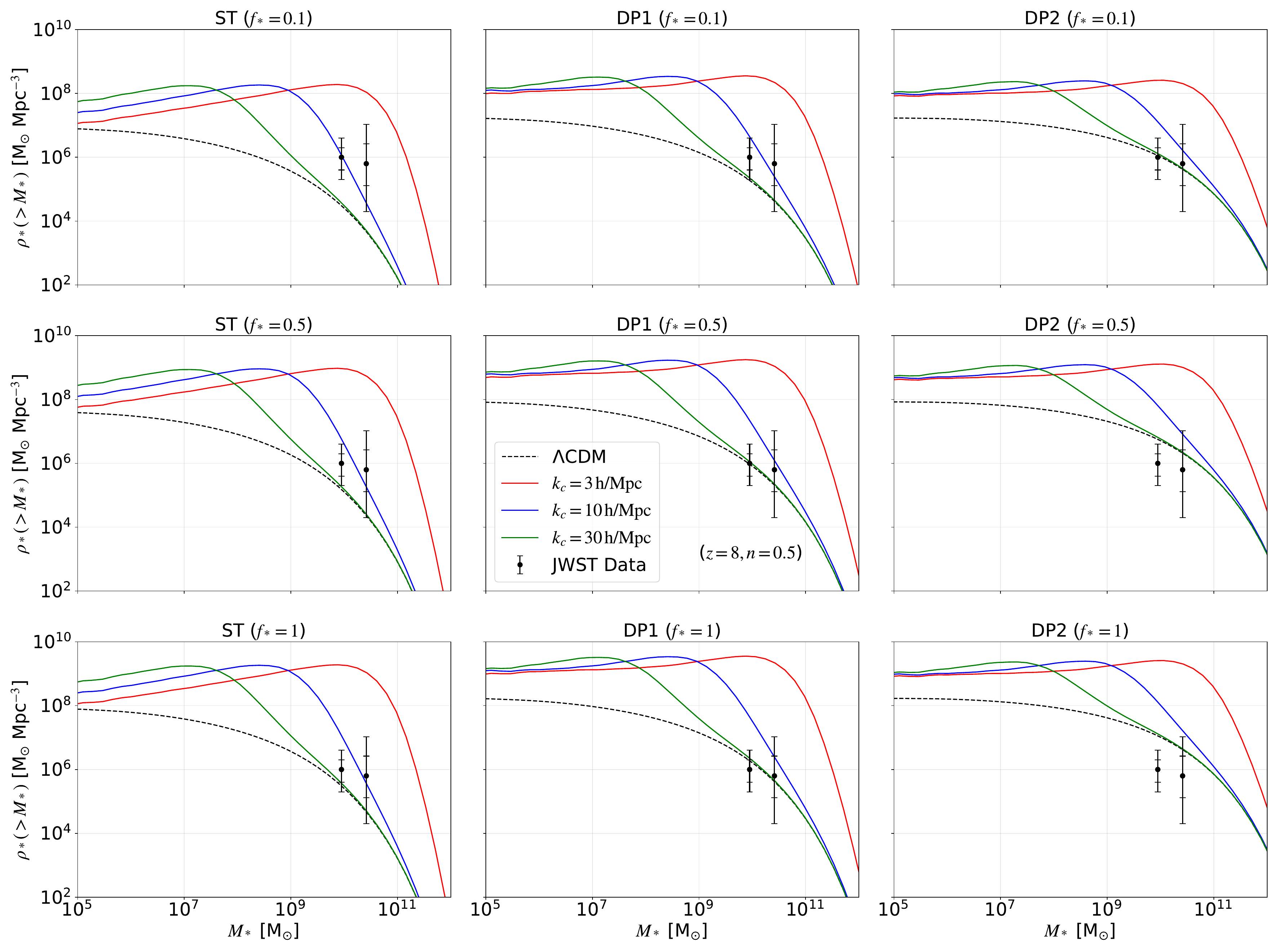}
\caption{Similar to Fig.\,\ref{Fig4a} but with $n=0.5$.}
\label{Fig4b}
\end{figure*}

\begin{figure*}
\centering
\includegraphics[width=1\linewidth]{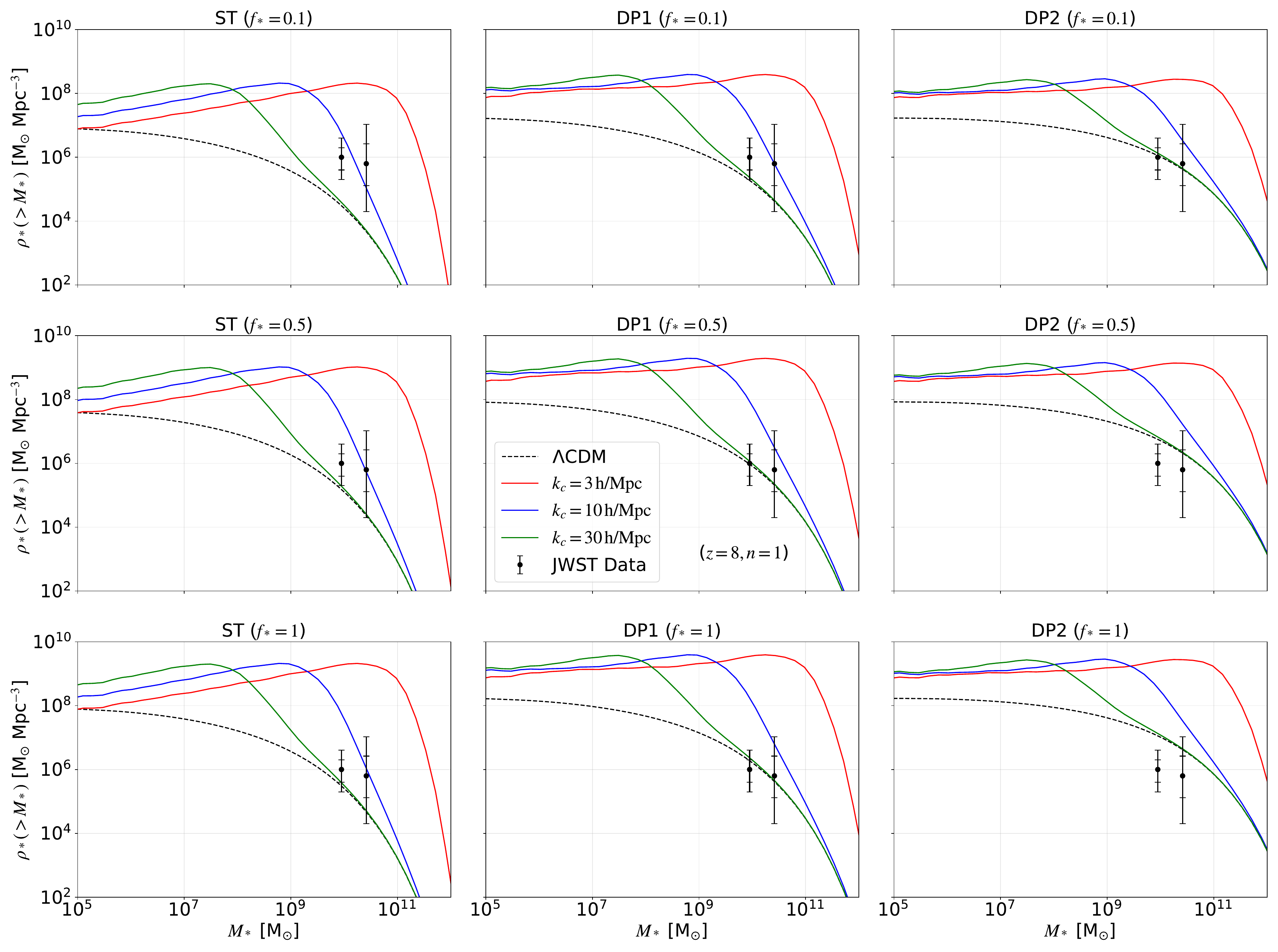}  
\caption{Similar to Fig.\,\ref{Fig4a} but with $n=1$.}
\label{Fig4c}
\end{figure*}

\section*{Appendix}\label{sec:a}
In this section, we present a complementary set of results that further support and enrich the main analysis, offering additional insights while preserving the clarity of the main text.
\subsection*{Enhanced Spectral Indices at $z \simeq 10$}
In Fig.\,\ref{Fig3b}, we have shown the cumulative stellar mass density evaluated at redshift $z = 10$ using three different halo mass functions, namely, ST, DP1, and DP2, under the assumption of an enhanced spectral index, $n = 0.5$. All other model parameters and assumptions are kept consistent with those employed in the previous analysis corresponding to Fig.\,\ref{Fig3a}. It is evident that varying the spectral index has no impact on the mass function predictions when considered within the context of the standard $\Lambda$CDM framework. The enhancement of the spectral index to $n = 0.5$ demonstrates a systematic amplification in the cumulative stellar mass density predictions across all three halo mass function models compared to the flat power spectrum case $n = 0$ within the context of the modified power spectrum. This enhancement is particularly pronounced for $M_\star > 10^7 M_{\odot}$ where massive galaxies dominate the stellar mass budget. 

On the other hand, adopting a spectral index of $n=0.5$ implies a steeper rise in the power spectrum at smaller scales, a behavior that aligns with certain theoretical scenarios for modified power spectra discussed in Sec.\,\ref{sec:ii}. The modified power spectrum with $n = 0.5$ effectively boosts the abundance of massive dark matter halos through enhanced small-scale clustering, enabling more efficient early stellar mass assembly that may bring theoretical predictions closer to the observed stellar mass densities (see Fig.\,\ref{Fig2}). Consequently, one anticipates a significant rise in the cumulative stellar mass density, particularly in the high-mass tails, as illustrated in Fig.\,\ref{Fig3b}. In line with this, compared to the results obtained for $n=0$, the cumulative stellar mass density predictions based on the ST, DP1, and DP2 formalisms for $n = 0.5$ tend to exhibit better agreement with the JWST observations when higher values of the characteristic scale are considered.

Our analysis for $n=0.5$ reveals an evident dependence of theoretical model predictions on both the adopted star formation efficiency and the characteristic scale under consideration. At low star formation efficiencies ($f_\star = 0.1$), one can find that different theoretical frameworks exhibit optimal agreement with JWST observations across distinct scales. In this regard, the ST formalism demonstrates consistency within the region of $3\,h\,\mathrm{Mpc}^{-1}\leq k_c\leq 10\,h\,\mathrm{Mpc}^{-1}$, the DP1 mass function shows concordance for $k_c\simeq 10\,h\,\mathrm{Mpc}^{-1}$, while the DP2 formulation yields predictions consistent with observations for $10\,h\,\mathrm{Mpc}^{-1}\leq k_c\leq 30\,h\,\mathrm{Mpc}^{-1}$.

On the other hand, under conditions of enhanced star formation efficiency (moderate to high $f_\star$ values), we observe a systematic shift in the scale-dependent performance of these models. Specifically, the ST formalism maintains consistency with JWST-derived cumulative stellar mass densities primarily for $k_c\simeq 10\,h\,\mathrm{Mpc}^{-1}$, while the DP1 mass function extends its region of agreement to encompass the range $10\,h\,\mathrm{Mpc}^{-1}\leq k_c\leq 30\,h\,\mathrm{Mpc}^{-1}$. Notably, under these elevated efficiency conditions, the DP2 model systematically overestimates the observed cumulative stellar mass density relative to the JWST benchmark, indicating potential limitations in its applicability at higher star formation rates. These findings underscore the scale- and efficiency-dependent nature of theoretical model validation and highlight the importance of adopting appropriate theoretical frameworks based on the specific physical conditions and observational scales under investigation.

The adoption of a steeper spectral index $n = 1$ produces the most dramatic enhancement in cumulative stellar mass density predictions. Following the methodology employed in our analysis of Figs.\,\ref{Fig3a} and \ref{Fig3b}, we have presented in Fig.\,\ref{Fig3c} the analogous results obtained when implementing the upper theoretical bound of the spectral index at $n=1$. Our findings demonstrate that even under this maximum spectral slope configuration, the modified power spectrum continues to exhibit amplification in the small-scale regime, resulting in a corresponding enhancement of the cumulative stellar mass density relative to the baseline scenarios examined at $n=0$ and $n=0.5$. Such enhancement demonstrates the exponential sensitivity of massive halo formation to small-scale power enhancements, where the modified matter power spectrum with $n = 1$ significantly boosts the collapse probability of rare, high-density peaks that correspond to massive dark matter halos at high redshift. 

Given these results, one can find out that the minimum systematic deviations between theoretical predictions from the ST, DP1, and DP2 mass functions and the observational constraints imposed by JWST data will necessitate the adoption of elevated characteristic scale parameters. However, the results corresponding to $n=1$ exhibit marginal deviations from those derived under the $n=0.5$ parameterization, with only select instances demonstrating modest alterations in the discrepancy between theoretical model predictions and observational constraints from JWST.

\subsection*{Enhanced Spectral Indices at $z \simeq 8$}

In Fig.\,\ref{Fig4b}, we have depicted the cumulative stellar mass density evaluated at redshift $z = 8$ using enhanced spectral index $n = 0.5$, following the same methodology as in the $z = 10$ analysis. The general trends observed are qualitatively similar to those obtained in the previous section, with systematic amplification in cumulative stellar mass density predictions across all three halo mass function models compared to the $n = 0$ case. However, several distinctive features emerge at $z = 8$ that differ from the $z = 10$ results.

The cumulative stellar mass density predictions at $z=8$ for enhanced spectral indices confirm the amplifying impact of increasing $n$ on reconciling theoretical models with JWST data. As shown in Fig.\,\ref{Fig4b} (for $n = 0.5$), all three halo mass function formalisms experience a systematic elevation in predicted mass densities relative to their $n = 0$ counterparts. This trend holds across various $k_c$ values and star formation efficiencies. Remarkably, the ST model, which typically demonstrates systematic underprediction relative to observations in the standard $\Lambda$CDM framework, attains observational concordance within the $2\sigma$ confidence interval for one scenario (while providing an exact fit in another case) when $f_\star = 0.1$ and the power spectrum is augmented with $k_c=10\,h\,{\rm Mpc^{-1}}$. Furthermore, the preditions improve substantially for elevated values of $f_\star$, particularly at larger values of $k_c$.

The DP1 model at $n = 0.5$ continues to display improved compatibility with JWST data across a broad swath of parameter space. For $f_\star = 0.1$, the predictions remain within or very close to the $1\sigma$ region for all values of $k_c \simeq 10\,h\,\mathrm{Mpc}^{-1}$. As $f_\star$ increases, particularly at $f_\star = 0.5$, DP1 maintains strong agreement for intermediate to high $10\,h\,\mathrm{Mpc}^{-1}\leq k_c\leq 30\,h\,\mathrm{Mpc}^{-1}$, demonstrating that even modest small-scale enhancements can be cansistent with the star formation efficiency assumptions. This reinforces the viability of combining mild spectral enhancements with physically motivated halo collapse physics to explain the rapid assembly of massive galaxies at $z \simeq 8$.

The prediction of DP2 model, while broadly consistent with the above trends, shows more sensitivity to increasing $f_\star$ at $n = 0.5$. At low efficiencies, its predictions fall comfortably within the $1\sigma$ bounds, particularly for larger $k_c$, reflecting the high-mass tail amplification. However, as $f_\star \rightarrow 1$, the DP2 curves tend to overshoot the observational bounds for certain $k_c$ values, echoing similar behavior seen in the $z \simeq 10$ case but now slightly more pronounced. This suggests that at $z \simeq 8$, the cumulative effects of enhanced clustering and increased cosmic time might push DP2 predictions into overproduction when paired with aggressive star formation scenarios.

In Fig.\,\ref{Fig4c}, which illustrates the case of $n = 1$ at redshift $z = 8$, the enhancement in the predicted stellar mass density becomes even more pronounced across all three mass function models. Notably, the ST formalism, typically considered the most conservative approach, achieves compatibility with JWST observations within the $1\sigma$ uncertainty band for low star formation efficiencies, particularly around $k_c = 10\, h\,\mathrm{Mpc^{-1}}$. Moreover, for $f_\star = 0.5$, the predictions fall within observational limits below the $1\sigma$ level, especially for $10\, h\,\mathrm{Mpc^{-1}} \leq k_c \leq 30\, h\,\mathrm{Mpc^{-1}}$, indicating a closer alignment with the data compared to the $n = 0$ case. This contrast highlights the exponential sensitivity of massive halo formation to small-scale enhancements in the power spectrum, particularly as the cosmic timeline allows for extended structure development.

The DP1 formalism under $n = 1$ exhibits an even more robust performance, delivering stellar mass density predictions that remain consistently aligned with JWST constraints across a wide range of star formation efficiencies and characteristic scales $10\, h\,\mathrm{Mpc^{-1}} \leq k_c \leq 30\, h\,\mathrm{Mpc^{-1}}$. This stability underscores the effectiveness of refined mass function prescriptions in modeling high-redshift galaxy formation, especially when informed by modifications in small- to intermediate-scale clustering. Crucially, the DP1 model achieves this alignment without artificially inflating the abundance of massive halos, thereby supporting its role as a balanced and physically grounded framework for early galaxy evolution.

On the other hand, the DP2 model demonstrates a tendency toward overprediction under the $n = 1$ scenario, particularly at maximal star formation efficiency ($f_\star = 1$) and at lower values of $k_c$. Although its predictions remain broadly within the $2\sigma$ observational range at intermediate scales, the model’s characteristic enhancement of massive halo populations, when combined with the steep spectral index and low star formation thresholds, yields outcomes that marginally remain observationally viable. This behavior suggests that, despite its efficacy in certain contexts, the DP2 approach might benefit from more conservative parameter selections or the incorporation of additional feedback mechanisms to mitigate excessive stellar mass buildup at lower redshifts.


\bigskip
\bibliography{draft_ml}
\end{document}